\documentclass[pre,twocolumn]{revtex4}

\usepackage{amsmath}
\usepackage{epsfig}

\begin{document}

\title{Cubatic phase for tetrapods}

\author{Ronald Blaak}
\email{blaak@thphy.uni-duesseldorf.de}
\affiliation{Institut f\"ur Theoretische Physik II,
  Heinrich-Heine-Universit\"at, D-40225 D\"usseldorf, Germany}

\author{Bela M. Mulder}
\email{mulder@amolf.nl}

\author{Daan Frenkel}
\email{frenkel@amolf.nl}
\affiliation{FOM Institute for Atomic and Molecular Physics, Kruislaan
  407, 1098 SJ Amsterdam, The Netherlands}

\begin{abstract}
We investigate the phase behavior of tetrapods, hard non-convex
bodies formed by 4 rods connected under tetrahedral angles. We
predict that, depending on the relative lengths of the rods these
particles can form a uniaxial nematic phase, and more surprisingly
they can exhibit a cubatic phase, a special case of the biaxial
nematic phase. These predictions may be experimentally testable,
as experimental realizations of tetrapods have recently become
available.
\end{abstract}

\maketitle

\section{Introduction}
\label{Sec:intro}

The simplest liquid crystalline phase is the nematic. It is a
spatially homogeneous phase in which the orientations of the
non-spherical component particles, e.g. rod-like or disk-like
colloids, are distributed in an anisotropic fashion. More
precisely, they are oriented around a preferred axis yielding a
phase with macroscopic uniaxial optical anisotropy. However, this
represents only the simplest form of rotational symmetry breaking.
When in addition the cylindrical symmetry around the nematic
director is broken, the phase that results is the so-called
biaxial nematic phase. As the name biaxial suggests, there are now
two preferred axes, which are mutually perpendicular. Biaxial
phases can be expected if the constituent particles themselves do
not have (effective) cylindrical symmetry, but are only invariant
under a limited number of discrete rotations
\cite{Freiser:1970PRL,Mulder:1989PRA}. Biaxial phase may also
appear in mixtures of rod-like and disk-like particles. Each of
the two components individually will form  uniaxial nematic phases
at sufficiently high densities. When mixed they will do the same,
but their mutual interaction is such that the preferred
orientation axis for the rods is perpendicular to that of the
disks
\cite{Alben:1973JCP,Stroobants:1984JPC,Camp:1997JCP2,Kooij:2000PRL}.

This still does not exhaust all possibilities for spatially
homogenous liquid-crystalline phases. Frenkel \cite{Frenkel:1991}
proposed that particles consisting of  three identical rods,
connected at right angles at their center, should form a stable
high-density phase with cubic orientational symmetry. This liquid
crystalline phase is referred to as a cubatic phase. It is a
special case of the general biaxial phase, since there are now
three mutually perpendicular axes of symmetry that are equivalent.

This model has subsequently been generalized to cross-like
particles ("Onsager crosses"), in which the three rods can
have unequal lengths \cite{Blaak:1998PRE}. These particles show a
surprisingly rich phase behavior. Not only do they form a cubatic
phase in the case that the three rods have approximately equal
lengths, but they can also show rod-like and plate-like behavior
by forming uniaxial nematic phases if one, respectively two, rods
are dominant in determining the shape of the particle. Moreover,
at higher densities, these uniaxial nematic phases become unstable
and different types of biaxial nematic phases are formed.

Unfortunately these lower-symmetry liquid-crystalline phases have,
thus far, not been observed in experiment. The main problem seems
to be that, unlike the rod-like colloidal particles that form
nematics, cross-like particles that are both rigid and
sufficiently monodisperse could not be made in a sufficient
quantities to allow a systematic study of their phase behavior. In
particular the cubatic phase has not yet been observed in
experiments, although simulations have suggested that a phase with
this symmetry may exist in a system of disk-like particles
\cite{Veerman:1992PRA}.

Recently, however, Alivisatos {\em et al.} have reported the
synthesis of colloidal CdTe tetrapods~\cite{Manna:2003NATM}. These
particles could be made with a high yield and with well-controlled
nanoscale dimensions. The experimental tetrapods consists of a
small crystalline body from which four arms grow under tetrahedral
angles. Since these arms are also crystalline, the tetrapods are
fairly rigid and, with a suitably chosen solvent (and proper
steric stabilization), should behave as rigid hard-core particles.

In this paper we consider the liquid crystalline behavior of
tetrapods. For simplicity we  work in the Onsager limit of large
aspect ratios and only take into account a hard-core interaction.
We  assume that the particles are monodisperse, but we treat all
possible combinations of relative lengths for the arms of the
tetrapod. We focus here on a bifurcation analysis, which gives us
an upper limit to the stability of the isotropic phase and yields
an indication of the nature of the more stable liquid crystalline
phases. Using this analysis, we argue that the tetrapods of
ref.~\cite{Manna:2003NATM} should, under certain conditions, form
cubatic phases.

In section \ref{Sec:model} we justify the main assumptions of the
model and derive an expression for the Helmholtz free energy. We make
use of rotation matrix elements, of which the main properties and
conventions are briefly described in the appendix. In section
\ref{Sec:bif} we perform the stability analysis of the isotropic phase
and interpret the results, and conclude with a discussion of the main
results in section \ref{Sec:dis}.

\section{The model}
\label{Sec:model}

To analyze the phase behavior of hard tetrapods, we need
expression for the free energy of this system. In general, this is
an intractable problem. However, for tetrapods with sufficiently
slender arms we can make the same assumptions that were introduced
by Onsager in the context of the isotropic-nematic transition of
thin hard rods~\cite{Onsager:1949AAS}. Onsager showed that for a
fluid of particles with large (strictly speaking, infinite)
length-to-width ratio, the excess free energy can be truncated at
the second virial coefficient level. In the case of hard-core
interactions this is equivalent to assuming that if one randomly
places particles with a given density in space, the probability
that three particles mutually overlap is negligibly small.

Although we have assumed large aspect ratios for the arms that
constitute a tetrapod, it is not immediately obvious that the
second virial approximation is valid. However, since tetrapods are
essentially objects with an open structure and consist of four
connected rod-like particles, one would expect that if two particles
overlap with each other this is mainly due to a single arm of one
particle that overlaps with a single arm of the other particle.

The validity of this plausibility argument is confirmed explicitely 
for Onsager crosses with three equally long arms in the isotropic
phase~\cite{Blaak:1998MP1}. In a detailed analysis it is shown that  
for a length-to-width ratio of about one thousand the probability that,
under the constraint that two particles overlap, more than a single
pair of the arms are overlapping, is less then a percent and decreases for
increasing aspect ratios. In other words, in the limit of large aspect
ratios one can describe the particle-particle interaction in terms of
independent pairs of arm-arm interactions only.

What Onsager showed for elongated particles in the isotropic phase is
that the asymptotic limit of the third virial coefficient $B_3$ can be
expressed in terms of the second virial coefficient and the aspect
ratio by 
$B_3 = B_2^2 {\cal O}(D/L\log(L/D))$~\cite{Onsager:1949AAS}. This has 
been confirmed by the calculation of virial coefficients for long
spherocylinders~\cite{Frenkel:1987JPC}. The fourth and fifth virial
showed a similar dependence $B_n = B_2^{(n-1)} {\cal O}(D/L)$. Since
the interaction between Onsager crosses in leading order is determined
by single rod-rod interactions the same behavior should be observed
for Onsager crosses, as indeed is found, provided aspect ratios are of
the order thousand~\cite{Blaak:1998MP1}. In addition, the main 
contribution to the higher virial coefficients stems from the
so-called ring diagrams, which would lead to same scaling behavior for
higher order virial coefficients. Hence corrections to the free
energy due to the simultaneous interaction of three or more particles
is an order $D/L$ smaller than the second virial contribution
and can therefore be neglected in the limit of infinite aspect
ratios. 

In summary then, the assumptions that virial expansion  of 
the free energy can be truncated at the second-virial level and that the 
interaction between particles can be considered as a sum of pair 
interactions between the "arms" of the particles, become exact in the 
limit of infinite aspect ratios.  For large but finite aspect ratios, 
these assumptions should constitute excellent approximations

The truncation of the virial series leads to the following free-energy
functional for homogeneous systems
\begin{equation}
\begin{split}
\label{Eq:free_func}
\beta f[\psi ] & = \beta \hat{f} + \int d\Omega \psi(\Omega) \ln
\psi(\Omega) \\
& +
\frac{1}{2} \rho \int d\Omega_1 \int d\Omega_2 \psi(\Omega_1)
\psi(\Omega_2) {\cal E} (\Omega_1,\Omega_2)
\end{split}
\end{equation}
Here $f$ is the Helmholtz free energy per particle, which is a
functional of $\psi$ the orientational distribution function
(ODF). This ODF is a measure for the fraction of particles with an
orientation $\Omega$, which is shorthand for the three Euler
angles $(\alpha,\beta\gamma)$ required to specify an arbitrary
orientation in a fixed reference frame, and is normalized to
unity. $\beta=(k_B T)^{-1}$ is the inverse temperature, $\rho$ the
number density and $\hat{f}$ the ideal gas term that does not
explicitly depend on the ODF. The second term corresponds to the
orientational entropy, while the third term is associated to the
translational entropy through the kernel ${\cal E}
(\Omega_1,\Omega_2)$ describing the over space integrated
interaction of two tetrapods with orientations $\Omega_1$ and
$\Omega_2$.

The interaction of hard-core objects is taken into account via the
excluded volume ${\cal E} (\Omega_1,\Omega_2)$. This volume, is
defined as the volume around particle $1$ (with orientation
orientations $\Omega_1$) that is inaccessible to particle $2$
(with orientation orientations $\Omega_2$). For two slender
particles, with lengths $L_1$ and $L_2$ and diameters $D_1$ and
$D_2$ respectively the excluded volume is, to leading order, given
by
\begin{equation}
\label{Eq:excl_vol_rods}
L_1 L_2 (D_1 + D_2) | \sin \gamma|,
\end{equation}
where $\gamma$ is the angle between the long axes of the
particles. Corrections to this expression are of order  $D/L$.
Since we are mainly interested in the limit of large aspect
ratios, we restrict ourselves to the leading order only.

In the case of tetrapods the excluded volume is, of course, more
complicated. First of all we now have four rod-like arms. We will
assume that the arms can be approximated by cylinders with
identical diameter $D$, but possibly different lengths $L_i$,
where $i=1, 2, 3, 4$. With the assumption that the arms of
tetrapods, whilst connected, interact independently, the excluded
volume of two tetrapods becomes a sum over the pairwise excluded
volumes of the arms
\begin{equation}
\label{Eq:excl_vol_tetra}
{\cal E} (\Omega_1,\Omega_2) = \sum_{i,j} 2 L_i^{(1)} L_j^{(2)} D |
\sin \gamma_{ij}^{12}|,
\end{equation}
where the superscript refers to the particle and the subscript to
the arm of the tetrapod, hence $\gamma_{ij}^{12}$ is the angle
between the $i$th and $j$th arm of particle 1 and 2 respectively.
For the isotropic phase this leads to a simple expression for the
second virial coefficient $B_2$. It is equal to half the excluded
volume averaged over all orientations, as it is simply equal to
the sum of second virials for all pairs of rods
\begin{equation}
\label{Eq:B2}
B_2 = \frac{\pi}{4} D \left( \sum_i L_i \right)^2.
\end{equation}

We note that we can expand any ODF depending on $\Omega$ in terms
of a linear combination of Wigner rotation matrix elements ${\cal
D}^l_{m,n}(\Omega)$. In what follows, we adopt the convention used
by Brink and Satchler in the description of the Wigner
matrices~\cite{Book:Brink-Satchler}.

In order to make use of the free energy functional
(\ref{Eq:free_func}), we  need to rewrite the interaction
(\ref{Eq:excl_vol_tetra}) in terms of these functions. To this
end, we introduce an arbitrary reference orientation of a
tetrapod. We  denote the directions of the arms with length $L_i$
of the reference tetrapod by the unit-vectors $\hat{e}_i$. For
simplicity we assume that $\hat{e}_1 = \hat{z}$ is the positive
$z$-direction and $\hat{e}_2$ is lies in the $xz$-plane with a
positive $x$-component. Since by definition the mutual directions
are under tetrahedral angles this fixes all directions.

This allows us to interpret the orientation $\Omega$ of a
particle, as the one we would obtain if we take the reference
particle and rotate it over the Euler angles denoted by $\Omega$.
Additionally we can also interpret $\Omega$ as the actual rotation
matrix, so the directions of the arms of a particle become $\Omega
\hat{e}_i$. Finally, we introduce the rotations $g_i$, such that $
\hat{e}_i \equiv g_i \hat{z}$. Note that these rotations $g_i$ are
not uniquely defined, since only two of the Euler angles are
required in order to satisfy the restriction. However,  this has
no effect on the final result.

It is obvious that the excluded volume (\ref{Eq:excl_vol_tetra})
of two tetrapods cannot depend on both orientations independently,
but only  on the relative orientation $\Omega_1^{-1} \Omega_2$.
Hence we can rewrite the excluded volume as
\begin{equation}
\label{Eq:excl_vol_tetra2}
\begin{split}
 & {\cal E} (\Omega_1^{-1} \Omega_2) \equiv {\cal E}
 (\Omega_1,\Omega_2) \\
 & = \sum_{i,j} 2 L_i^{(1)} L_j^{(2)} D | \Omega_1 \hat{e}_i^{(1)}
 \times \Omega_2 \hat{e}_j^{(2)}| \\
 & = \sum_{i,j} 2 L_i^{(1)} L_j^{(2)} D |\hat{z}^{(1)} \times
g_i^{-1} \Omega_1^{-1} \Omega_2 g_j \hat{z}^{(2)}|.
\end{split}
\end{equation}

With the aid of this form we can now expand the excluded volume
in terms of rotation matrix elements ${\cal D}^l_{m,n}(\Omega)$, by
introducing the expansion coefficients $E_{l,m,n}$
\begin{equation}
\label{Eq:excl_vol_exp}
{\cal E} (\Omega) \equiv \sum_{l,m,n} E_{l,m,n} {\cal D}^l_{m,n}(\Omega),
\end{equation}
where $l=0,1,\cdots,\infty$, and $-l \leq m,n \leq l$.
Strictly speaking, the rotation matrix elements also defined for 
half-integer "spin" values. However, for reasons of symmetry these 
can be omitted~\cite{Book:Brink-Satchler}. 

The expansion coefficients can be evaluated by using the orthogonality
relation (\ref{Eq:D-ortho}) for the rotation matrix elements
\begin{equation}
\label{Eq:excl_vol_coef}
E_{l,m,n} = \frac{2 l + 1}{8 \pi^2} \int d \Omega
{\cal D}^{l~*}_{m,n}(\Omega) {\cal E} (\Omega).
\end{equation}
Substituting the expression~(\ref{Eq:excl_vol_tetra2}) and changing
the integration variables we get
\begin{widetext}
\begin{equation}
\begin{split}
\label{Eq:excl_vol_coef2}
E_{l,m,n}
& = \frac{2 l + 1}{8 \pi^2} 2 D \sum_{i,j} L_i L_j \int d
\Omega {\cal D}^{l~*}_{m,n}(g_i \Omega g_j^{-1}) |\hat{z}^{(1)} \times
\Omega \hat{z}^{(2)}| \\
& = \frac{2 l + 1}{8\pi^2} 2 D \sum_{p,q}
\left( \sum_i L_i {\cal D}^{l~*}_{m,p}(g_i)\right)
\left( \sum_j L_j {\cal D}^{l  }_{n,q}(g_j)\right)
\int d \Omega {\cal D}^{l*}_{p,q}(\Omega) |\sin \beta|
\end{split}
\end{equation}
\end{widetext}
where we made use of the  symmetry relation (\ref{Eq:D-sym}) and
closure relation (\ref{Eq:D-closure}) and replaced the cross product
by its representation in Euler angles $|\sin \beta|$.

In order for the integral to be non-zero, it is required that both
indices of the rotation matrix element are zero. This is a special
case for which the function reduces to a Legendre polynomial
${\cal
  D}^{l}_{0,0}(\Omega) = P_l(\cos(\beta))$. By introducing the
following shorthand notations
\begin{eqnarray}
E_{l,m} & \equiv & \sum_i L_i {\cal D}^{l}_{m,0}(g_i) \\
\mu_l & \equiv & \frac{2 l+ 1}{2} \int_0^\pi d \beta P_l(\cos \beta)
\sin^2 \beta
\end{eqnarray}
the expansion coefficients of the excluded volume can be written in a
compact form as
\begin{equation}
\label{Eq:excl_vol_coef3}
E_{l,m,n} = (2 D) \mu_l E_{l,m}^* E_{l,n}.
\end{equation}
The integral that remains can be readily evaluated (See
\cite{Book:Gradshteyn-Ryzhik} Eq. 7.132.1) and is only non-zero for even value
of $l$
\begin{equation}
\begin{split}
\mu_{2 l} & =  - \frac{\pi (4 l + 1)}{(l+1)(2l-1)2^{4l+2}}  \binom{2l}{l}^2 \\
\mu_{2 l+1} & = 0.
\end{split}
\end{equation}

Finally we introduce another shorthand notation by using the kernel
(\ref{Eq:excl_vol_tetra}) as a functional acting on an arbitrary
function $\psi(\Omega)$
\begin{equation}
\label{Eq:excl-func}
{\cal E}[\psi](\Omega) \equiv \int d \Omega' {\cal E}(\Omega'^{-1}
\Omega) \psi(\Omega').
\end{equation}
In particular we allow it to operate on a rotation matrix
element. Using the expansion (\ref{Eq:excl_vol_exp}) and the properties
(\ref{Eq:D-sym}) and (\ref{Eq:D-closure}) this can be manipulated to
yield
\begin{equation}
\label{Eq:excl-func-D}
{\cal E}[ {\cal D}_{m,n}^l ] = \sum_p E_{l,n,p} {\cal D}_{m,p}^l(\Omega).
\end{equation}
Note that this generates a linear combination of rotation matrix
elements with the same value for $l$ and $m$. In other words each
set ${\cal D}_{m,n}^l$ with $n=-l,\cdots,l$ forms a subset of
functions that is invariant under the functional operator of the
excluded volume. We can go one step further by evaluating
eigenfunctions of the excluded volume. Using the special form of
the coefficients (\ref{Eq:excl_vol_coef3}), one can easily check
that for each combination of $l$ and $m$ at most a single
eigenvector $\chi^l_{m}$ exist with a non-zero eigenvalue
$\lambda_l$
\begin{equation}
\label{Eq:chi}
\chi^l_{m}(\Omega) = \sum_p E_{l,p} {\cal D}_{m,p}^l(\Omega)
\end{equation}
\begin{equation}
\label{Eq:lambda}
\lambda_{l} = (2 D) \mu_l \sum_p E_{l,p}^* E_{l,p}.
\end{equation}
Note that the eigenvalue is independent of $m$ and in special cases
also might become zero as for instance for odd values of $l$.

\section{Bifurcation Analysis}
\label{Sec:bif}

The thermodynamically stable phase of our model, is described by
the ODF that minimizes the free energy (\ref{Eq:free_func}).
Usually this free energy is not known exactly and one uses a
truncated expansion as an approximate function. There is however
one exception: the isotropic phase. In the limit of infinite
dilution, particles do not interact and therefore each orientation
has the same probability, hence the ODF is merely a constant.

With the aid of a bifurcation analysis we can determine an upper
limit to the stability of the isotropic phase. To this end, we
make an expansion of the ODF around the stable isotropic solution.
Rather than inserting this into the free energy
(\ref{Eq:free_func}), we use this to find solutions of the
stability equation that is obtained as the functional derivative
of the free energy with respect to the ODF
\begin{equation}
\label{Eq:free-stable}
\frac{\delta }{\delta \psi(\Omega)} \left\{ \beta f[\psi] - \lambda
\int d \Omega \psi(\Omega) \right\} = 0,
\end{equation}
where $\lambda$ is a Lagrange multiplier to take care of the proper
normalization of the ODF. Evaluating this expression and using the
definition (\ref{Eq:excl-func}) this gives us
\begin{equation}
\label{Eq:free-stable2}
\ln(\psi(\Omega)) + \rho {\cal E}[\psi](\Omega) = \lambda.
\end{equation}

For both the ODF $\psi$ and number density $\rho$ we take the formal
expansion in a small parameter $\epsilon$
\begin{eqnarray}
\label{Eq:psi-exp}
\psi & = \psi_0 + \epsilon \psi_1 + \epsilon^2 \psi_2 + \cdots \\
\label{Eq:rho-exp}
\rho & = \rho_0 + \epsilon \rho_1 + \epsilon^2 \rho_2 + \cdots.
\end{eqnarray}
Here we use $\psi_0 = \frac{1}{8 \pi^2}$ as the ODF for the isotropic
phase. By inserting these expansions in the stability equation
(\ref{Eq:free-stable2}) and grouping terms for each power in
$\epsilon$, we obtain the bifurcation equations. By solving these we
can find the lowest density $\rho_0$ at which a
symmetry breaking mode exists that lead to a lower free energy than
that of the isotropic phase. In the case of a non-zero value for
$\rho_1$ this is sign of a first order phase transition at a lower
density and hence $\rho_0$ is the upper limit for the meta-stability of the
isotropic phase.

The zeroth-order bifurcation equation merely states that the
isotropic solution $\psi = \psi_0$ is a solution of the
stationarity equation. The first order bifurcation has the form of
an eigenvalue problem
\begin{equation}
\label{Eq:bif_eq1}
\frac{\psi_1}{\psi_0} + \rho_0 {\cal E}[\psi_1] = 0,
\end{equation}
where we have already eliminated the constant contributions. Since we
are interested in a non-trivial solution that leads to the lowest
possible positive value of $\rho_0$, we only need to consider linear
combinations of eigenfunctions (\ref{Eq:chi}). In particular we need
to find the one that has the largest absolute value among all negative
eigenvalues.

One can show that, for the case of tetrapods, there are only two
eigenvalues that can fulfill that requirement, namely the ones
corresponding to $l=2$ and $l=4$
\begin{equation}
\label{Eq:lambda2}
\lambda_2 = - \frac{\pi^3}{6} D L^2 (4 R^2 - 1)
\end{equation}
\begin{equation}
\label{Eq:lambda4}
\lambda_4 = - \frac{\pi^3}{1296} D L^2 (80 R^2 + 1).
\end{equation}
For practical purposes we used here $L=\sum_i L_i$, and $R^2 = (\sum_i
L_i^2)/L^2$. Since $\lambda_2 = \lambda_4$ for $R^2=\frac{31}{112}$ and by
construction $\frac{1}{4} \leq R^2 \leq 1$, we need to distinguish two
types of tetrapods. The ones with $R^2 > \frac{31}{112}$ for which we
need to consider modes related to $l=2$ and the ones with $R^2 <
\frac{31}{112}$ for which the modes with $l=4$ are the important
ones. For the special case $R^2=\frac{31}{112}$ we would actually need
to consider combinations of both, which makes the analysis somewhat more
involved, but since this is not going to lead to new insights we will
not treat it separately.

\begin{figure}[t]
\epsfig{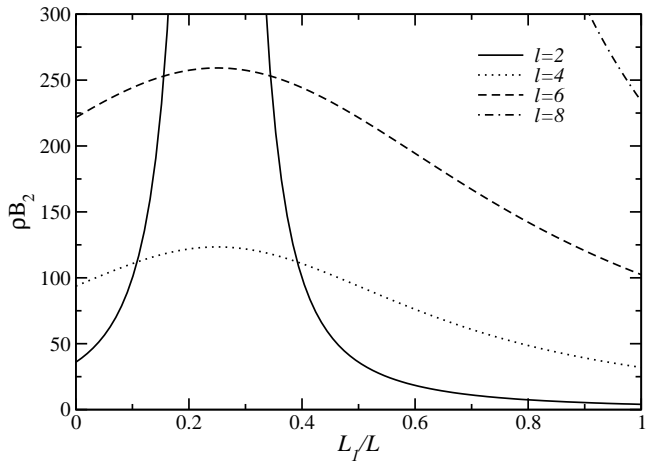}
\caption[a]{
  The density at which the isotropic phase becomes unstable with
  respect to different modes characterized by $l$ for a subset of
  particles with equal volume and arm lengths
  $L_2=L_3=L_4=(L-L_1)/3$. The solid line ($l=2$) denotes the
  isotropic-nematic instability and the dotted line ($l=4$) indicates
  the isotropic-cubatic instability.}
\label{Fig:bif}
\end{figure}

As an illustration we show in Fig.~\ref{Fig:bif} the location of the
four lowest instabilities of the isotropic phase for a specific class
of particles, i.e. those for which $L_2=L_3=L_4=(L-L_1)/3$. This set
includes the fully symmetric tetrapod ($L_1/L=1/4$) and a limiting
tetrapod with only a single arm ($L_1/L = 1$). Note that the volume of
each particle is the same $v=(\pi/4) D^2 L$, but the density at
which the isotropic phase becomes unstable for the symmetric tetrapod
is approximately thirty times higher than that of a tetrapod with a
single arm. The volume fraction $\phi = \rho v$, however, is
proportional to $D/L$, which means that in the limit of large aspect
ratios the transition takes place at small volume fractions.

Although we now know the upper limit to the stability of the
isotropic phase, we do not yet know which are the symmetry
breaking modes. To find these,  we need to perform a second order
bifurcation analysis, i.e. solve the equation
\begin{equation}
\label{Eq:bif_eq2}
\frac{\psi_2}{\psi_0} - \frac{1}{2} \left( \frac{\psi_2}{\psi_0}
\right)^2 + \rho_0 {\cal E}[\psi_2] + \rho_1 {\cal E}[\psi_1] = - \int d
\Omega \frac{\psi_1^2}{2 \psi_0^3}
\end{equation}
employing the general solution of the first order bifurcation
equation given by (\ref{Eq:bif_eq1})
\begin{equation}
\label{Eq:bif_eq1-sol}
\psi_1(\Omega) = \sum_m c_m \chi^l_m(\Omega),
\end{equation}
where the $c_m$ are some complex constants and $l$ is either 2 or
4. Substitution in the second-order bifurcation equation
(\ref{Eq:bif_eq2}), multiplying with $\chi^{l*}_n$ and integrating
over the orientation $\Omega$, gives us a set of coupled
non-linear equations in the coefficients $c_m$ and constant
$\rho_1$
\begin{equation}
\label{Eq:bif_eq2-sol2}
\rho_1 \lambda_l c_n \int d \Omega \chi^{l*}_n \chi^{l}_n
= \frac{1}{2 \psi_0^2} \int d \Omega \chi^{l*}_n\psi_1^2.
\end{equation}
We can also extract the value of $\rho_1$, if we use the complete
function $\psi_1^*$ in stead of $\chi^{l*}_n$
\begin{equation}
\label{Eq:bif_eq2-rho1}
\rho_1 = \frac{1}{2 \lambda_l \psi_0^2} \frac{\int d \Omega
  \psi_1^3}{\int d \Omega \psi_1^2}.
\end{equation}
It is important to realize that there is a restriction on
$\rho_1$. The reason is that a non-zero value of $\rho_1$ is
associated with a first order phase transition and that in order
to follow the solution towards lower densities we need a
non-positive value, hence $\rho_1 \leq 0$.

Let us now consider the case $R^2 > \frac{31}{112}$ with $l=2$ in
the preceding equations. The set of equations
(\ref{Eq:bif_eq2-sol2}) can be solved analytically and yields only
a single non-trivial solution, which is degenerate since all
rotations of a solution are also solutions of the set of
equations. For a conveniently chosen reference frame the solution
is
\begin{equation}
\label{Eq:bif_eq1-psi1-2}
\psi_1(\Omega) = c_0 \chi^2_0(\Omega)
\end{equation}
It bifurcates at a reduced density
\begin{equation}
\label{Eq:bif_eq1-rho0-2}
\rho_0 B_2 = \frac{12}{4 R^2 - 1}.
\end{equation}
Note that in the limit of a single arm ($R^2=1$) this reduces to
the correct result for uniaxial rod-like particles. This
particular solution is invariant under rotations about the
$z$-axis, and hence one can expect that the system shows a phase
transition from an isotropic to an uniaxial nematic phase.
Inserting the solution in the expression (\ref{Eq:bif_eq2-rho1})
for $\rho_1$ we obtain
\begin{equation}
\label{Eq:bif_eq1-rho1-2}
\begin{split}
\rho_1 = & \frac{- \pi^3 D c_0}{14 \psi_0^2 \lambda_2^2}
[L_1 + L_2 - L_3 - L_4] \times \\
& [L_1 - L_2 + L_3 - L_4] [L_1 - L_2 - L_3 + L_4].
\end{split}
\end{equation}
From this result it follows that if the sum of the lengths of two
arms equals the sum of the length of the two remaining arms we
find $\rho_1=0$. These special particle configurations could
therefore possibly lead to a continuous phase transition and be
the source of a biaxial phase. Whether this is scenario really
applies, cannot be determined from this analysis. One could
resolve this issue either by solving higher order bifurcation
equations or by a full numerical minimization of the free energy.
This falls outside the scope of the present paper. For other
particle configurations the requirement of non-positive values for
$\rho_1$ will fix the sign of the coefficient $c_0$, which can be
positive or negative, and hence fully determine the solution
(\ref{Eq:bif_eq1-rho0-2}), because the magnitude will only depend
on the choice of normalization.

Similar to the case for Onsager crosses \cite{Blaak:1998PRE}, we
can interpret this phenomenon in terms of rod-like and disk-like
behavior. For each of the four arms of the tetrapod, we can
determine the nematic order parameter, which is defined as the
average value of $\frac{3}{2} \cos^2(\theta)-\frac{1}{2}$ with
$\theta$ the angle between the nematic axis and the direction of
the arm. It can be shown that, up to a positive normalization
factor, this is proportional to $c_0 (4 L_i - L)/3$ for all arms.
Making use of permutations of arms it follows that for positive
values of $c_0$ the longest arm has the largest nematic order,
while for negative $c_0$ it would be the shortest arm. A special
limit of the former, is the case where $L_1$ is much larger than
the other three. It is obvious that for such particles this
longest arm will dominate the behavior and the tetrapod behaves as
a single rod-like particle. The other extreme occurs when $L_1$ is
much smaller than the other three. In that case the competition
among those three arms does not allow any of them to dominate and
there is a preference for them to be perpendicular to the nematic
axis and, as in the case of disks,  it is the smallest dimension
that determines the orientation  of the particle.

\begin{figure}[t]
\vspace{-1cm}
\epsfig{figure=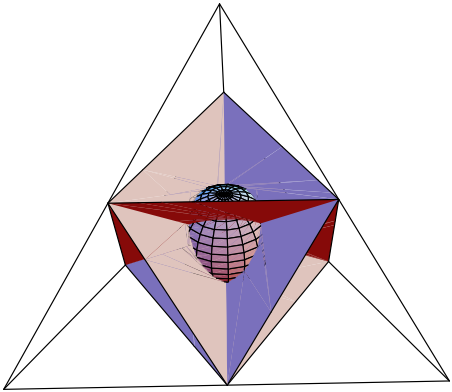,width=8.5cm,angle=0}
\vspace{-2cm}
\caption[a]{
  Phase diagram of the nature of symmetry breaking modes leading
  to the instability of the isotropic phase. Each point inside the
  tetrahedron corresponds to a given shape of the tetrapod. A vertex
  corresponds to a single arm, the opposite plane to a tetrapod with
  three arms. For particles inside the sphere a cubatic phase is
  expected, outside the sphere a nematic phase which is either
  rod-like, if it contains a vertex, or platelet-like if it does
  not. Both species are separated by planes denoting the particles
  that might have a continuous phase transition and show biaxial
  behavior.}
\label{Fig:bif3d}
\end{figure}

The set of equations (\ref{Eq:bif_eq2-sol2}) can also be solved
for the case of $R^2 < \frac{31}{112}$, but then we need to put
$l=4$ in the equations
(\ref{Eq:bif_eq1-sol})-(\ref{Eq:bif_eq2-rho1}). The bifurcation
density can easily be determined from the proper eigenvalue
(\ref{Eq:lambda4})
\begin{equation}
\label{Eq:bif_eq1-rho0-4}
\rho_0 B_2 = \frac{2592}{80 R^2 + 1}.
\end{equation}
But instead of having a single family of solutions that solve the
equations (\ref{Eq:bif_eq2-sol2}), we now have two. The first
family corresponds again to a uniaxial nematic phase for which
\begin{equation}
\label{Eq:bif_eq1-psi1-4N}
\psi_1(\Omega) = c_0 \chi^4_0(\Omega)
\end{equation}
is the particular solution invariant under rotations about the
$z$-axis, and $c_0<0$ in order for $\rho_1$ to be negative. Contrary
to the previous case $c_0$ does not change sign.

A particular solution of the second family of solutions is given by
\begin{equation}
\label{Eq:bif_eq1-psi1-4O}
\psi_1(\Omega) = c_0 \left\{ \chi^4_0(\Omega) + \frac{5}{14} \left(
  \chi^4_{4}(\Omega) + \chi^4_{-4}(\Omega) \right) \right\}.
\end{equation}
This solution is only invariant under discrete rotations over
$\pi/2$ about the $x$-, $y$-, and $z$-axes. It therefore
corresponds to the cubic symmetry group and hence to a cubatic
phase. Also in this case,  the constraint on $\rho_1$ results in a
negative value for $c_0$.

\begin{figure}[t]
\epsfig{figure=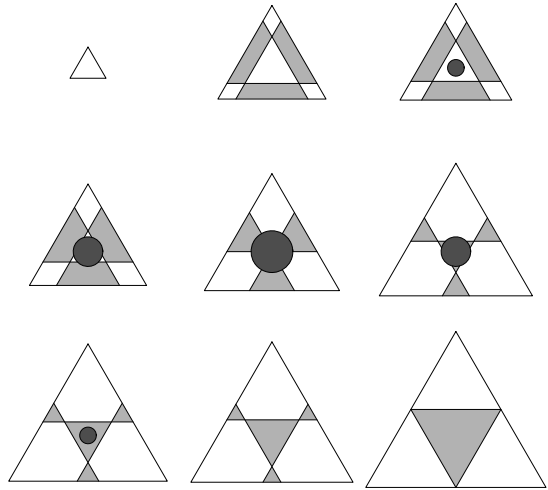,width=7.5cm,angle=0}
\caption[a]{
  Cross-sections of the phase diagram for constant value of $l_i$,
  from left to right, top to bottom for decreasing values. The dark
  circular areas correspond to the cubatic solution, the light and
  white areas to the two types of nematic solutions, being rod- and
  disk-like respectively.}
\label{Fig:bif2d}
\end{figure}

This analysis allows us to sketch a tentative phase diagram (see
 Figure \ref{Fig:bif3d}). In this
a schematic figure, we indicate the nature of the
symmetry-breaking modes that lead to the instability of the
isotropic phase. We can characterize the particle shape by the
normalized values $l_i = L_i/L$ with the constraint
$l_1+l_2+l_3+l_4=1$. Each possible particle shape, i.e.
combination of relative lengths of the arms, corresponds to a
point inside a tetrahedron, which is the projection of the
4-dimensional "shape" space. A vertex of the tetrahedron
represents the limit of a tetrapod with only a single arm, the
triangular plane opposite to the vertex contains all particle
shapes for which that same arm has zero length while the remaining
three arms have non-zero lengths. In Figure \ref{Fig:bif2d}, we
have made cross-sections corresponding to planes with constant
value for one of the $l_i$.

Nine distinct regions can be identified:  a spherical region
corresponding to $R^2 \leq \frac{31}{112}$, where there are two
modes that lead to the instability of the isotropic phase, one of
which yields the cubatic solution. In Figure \ref{Fig:bif2d} these
regions are indicated as the dark circular areas. Outside these
regions, we find the particles for which there is only a mode with
uniaxial symmetry that makes the isotropic phase unstable. There
are four equivalent areas close to a vertex, shown as white in
Figure \ref{Fig:bif2d}, that represent tetrapods where the
solution (\ref{Eq:bif_eq1-psi1-2}) has a positive sign, and four
equivalent areas, indicated by light gray in Figure
\ref{Fig:bif2d}, where the solution has a negative sign. They are
separated by the planes(lines) leading to ``biaxial'' particles.

\section{Discussion}
\label{Sec:dis} The present work suggests that the availability of
nanocrystalline tetrapods with a well controlled size and shape
~\cite{Manna:2003NATM}, may make it possible to observe cubatic
liquid crystalline phases in experiments. Of course, the present
"Onsager-style" analysis only becomes exact in the limit of very
slender, rigid arms.  Within this approximation, we have
determined the upper limit to the stability of the isotropic phase
by means of a bifurcation analysis. In addition, we have
determined the nature of the fluctuations that cause the
instability.

Roughly speaking we can distinguish two types of tetrapods, the
ones for which the arms have approximately the same lengths and
the ones where one or more arms are significantly longer than the
others. For the first group the isotropic phase becomes unstable
with respect to a distortion with either nematic or cubatic
symmetry, while for the second group only a nematic symmetry comes
into play.

In general the transitions will be first order since we obtained a
non-zero value of the first order shift in density $\rho_1$ along
the bifurcating solutions, indicating the presence of a v.d. Waals
loop. An exception might be formed by the particles located in the
planes in Figure \ref{Fig:bif3d}. Although the results of the
bifurcation analysis cannot guarantee that the symmetry of the
fluctuations that lead to the instability will also be the
symmetry of the more stable phase, the experience in a similar
study of Onsager crosses has shown that the bifurcation analysis
has a high predictive value \cite{Blaak:1998PRE}. In addition, it
also strongly indicates that a system of tetrapods with
approximately identical arms will have an isotropic to cubatic
phase transition, which in principle could be verified by a full
minimization of the Helmholtz free energy functional.

Based on the results of the Onsager crosses, we predict that at
higher densities a system of tetrapods will undergo additional
transitions to phases with yet lower symmetry, ultimately arriving
at the phase where only arms with identical lengths are aligned.
In particular three intermediate phases might appear that are
invariant under 2-fold, 4-fold, and 6-fold rotations from the
cubic group. In some cases these phases could actually preempt the
isotropic-to-nematic or isotropic-to-cubatic phase transition.
This is most likely to happen for particle shapes close to the
planes and/or surface that separate the different regions in the
phase diagram. A full numerical free-energy minimization would be
required to confirm the existence of these phases and to ascertain
whether a transition from isotropic phase to any of these four
phases is possible.

The formation of a cubatic phase even for fully symmetric
tetrapods may seem surprising. Naively, one might expect to
observe a "tetrahedratic" liquid crystalline phase. The reason why
the latter phase does not appear here is related to the fact that
in the Onsager approximation presented here the arms of a particle
effectively interact independently with those of other particles.
Essentially we are therefore insensitive to the details on how the
arms are connected. Although the relative orientation of the arms
within each particle is fully accounted for, we cannot specify
that the arms are connected end-on. The same results would
therefore be obtained for particles in which the four rods were
connected at their mid-points. Such particles would in fact have
cubic symmetry and are hence are unable to form a  phase with
tetrahedral symmetry.

The results presented here are only valid in the limit of infinite
aspect ratios. For finite aspect ratios we expect two types of
corrections: (i) contributions due to simultaneous overlap of
three or more arms and (ii) dependencies on the detailed
construction of the particles. In practice, one would presumably
require aspect ratios of the order one thousand or more in order
to make these corrections negligible (see Ref.\
\cite{Blaak:1998PRE}). Nevertheless one could expect that our
qualitative findings remain valid even for smaller aspect ratios,
as is the case for single rods. The fact that particles might not
be perfectly monodisperse is probably not a problem, since there
is a rather broad range of particle shapes that gives rise to the
cubatic instability of the isotropic phase. However, finite aspect
ratios imply that also non-homogeneous phases should be
considered, in which the isotropic-to-cubatic transition might be
preempted by crystallization. Finally, there may be kinetic
limitations to the formation of cubatic phases of tetrapods, as
the tetrapods are likely to become entangled at high densities,
which could lead to kinetically arrested glass-like phases.

\appendix
\section{Rotation matrix elements}
Here we list the main properties of the rotation matrix elements. For
a more extended discussion we refer the reader to
\cite{Book:Brink-Satchler}.

A rotation $\Omega = (\alpha,\beta,\gamma)$ is obtained by successive
rotations of angles $\alpha$, $\beta$, and $\gamma$ about the $z$-,
$y$-, and $z$-axis respectively. The invariant measure of the rotation
is given by $d \Omega = \sin(\beta) d \alpha d\beta d\gamma$, with
$\alpha,\gamma \in [0,2 \pi]$ and $\beta \in [0,\pi]$.

Symmetry relation
\begin{equation}
\label{Eq:D-sym}
{\cal D}^{l~*}_{n,m} (\Omega) = {\cal D}^l_{m,n} (\Omega^{-1})
\end{equation}

Orthogonality relation
\begin{equation}
\label{Eq:D-ortho}
\int d \Omega {\cal D}^{l'~*}_{m',n'}(\Omega) {\cal
  D}^{l}_{m,n}(\Omega) = \frac{8 \pi^2}{2 l + 1} \delta_{l,l'}
  \delta_{m,m'} \delta_{n,n'}
\end{equation}

Closure relation
\begin{equation}
\label{Eq:D-closure}
{\cal D}^l_{m,n} (\Omega_2 \Omega_1) = \sum_{p=-l}^l {\cal D}^l_{m,p}
(\Omega_2) {\cal D}^l_{p,n} (\Omega_1)
\end{equation}

\end{document}